\documentclass[aps,pra,amsmath,amssymb,superscriptaddresss,nofootinbib,twocolumn,showpacs]{revtex4}

\pdfoutput=1

\usepackage[bookmarks = true, pdfpagemode = None, pdfstartview = FitH, colorlinks = true, urlcolor = blue,hyperfootnotes=false]{hyperref}
\usepackage{graphicx}
\usepackage{pgf,pgfnodes,pgfarrows}

\newcommand{\leftexp}[2]{{\vphantom{#2}}^{#1}\!{#2}}
\newcommand{\ket}[1]{| #1 \rangle}
\newcommand{\bra}[1]{\langle #1 |}
\newcommand{\CZ}{\leftexp{C}{Z}}
\newcommand{\CX}{\leftexp{C}{X}}
\newcommand{\set}[1]{\mathcal{#1}}
\newcommand{\setH}{\set{H}}
\newcommand{\setS}{\set{S}}
\newcommand{\setZ}{\set{Z}}
\newcommand{\setM}{\set{M}}
\newcommand{\setMS}{\set{M}_{\mbox{\scriptsize{S}}}}
\newcommand{\smallsetMS}{\set{M}_{\mbox{\tiny{S}}}}
\newcommand{\setMH}{\set{M}_{\mbox{\scriptsize{H}}}}
\newcommand{\smallsetMH}{\set{M}_{\mbox{\tiny{H}}}}
\newcommand{\setMSE}{\set{M}_{\mbox{\scriptsize{SE}}}}
\newcommand{\smallsetMSE}{\set{M}_{\mbox{\tiny{SE}}}}

\begin{document}

\title{Graphical description of Pauli measurements on stabilizer states}

\author{Matthew B.~Elliott}
\email{mabellio@unm.edu} \affiliation{Department of Physics and
Astronomy, MSC07--4220, University of New Mexico, Albuquerque, NM
87131-0001}

\author{Bryan Eastin}
\affiliation{National Institute of Standards and Technology,
Boulder, CO 80305}

\author{Carlton M.~Caves}
\affiliation{Department of Physics and Astronomy, MSC07--4220,
University of New Mexico, Albuquerque, NM 87131-0001}
\affiliation{Department of Physics, University of Queensland, Brisbane,
QLD 4066, Australia}

\begin{abstract}

We use a graphical representation of stabilizer states to describe,
simply and efficiently, the effect of measurements of Pauli products
on stabilizer states.  This work complements our earlier work [Phys.
Rev.~A \textbf{77}, 042307 (2008)], which described in graphical
terms the action of Clifford operations on stabilizer states.

\end{abstract}


\maketitle

\section{Introduction}

Recently we introduced a graphical representation of stabilizer
states and translated the action of Clifford operations on stabilizer
states into graph operations on stabilizer-state
graphs~\cite{elliott:graphs}. The purpose of that paper was, in part,
to augment the stabilizer formalism by providing techniques for
understanding and manipulating this important class of states.  The
purpose of this paper is to extend our previous results by describing
graphically the effect of measurements of Pauli products on
stabilizer states~\cite{hein:entanglement}.\footnote{See also
Ref.~\cite{schlingemann:cluster} for a single-qubit measurement rule
applied to a very different graphical representation of stabilizer
states.}

Pauli measurements are the natural set of measurements to consider in
the context of stabilizer states, because the post-measurement state,
as an eigenstate of the measured Pauli product, is also a stabilizer
state.  Since we can represent both the pre- and post-measurement
states by graphs, the effect of the measurement can be represented by
a graph transformation.

Section~\ref{sec:background} reviews the concept of stabilizer-state
graphs and lists some relevant results from
Ref.~\cite{elliott:graphs}.  This is not meant to be a complete
introduction to stabilizer-state graphs and is surely insufficient
background to enable their comfortable manipulation.  Nonetheless,
anyone comfortable with stabilizer states and quantum circuits should
find the review here sufficient for the needs of this paper and can
consult Ref.~\cite{elliott:graphs} and references therein for further
details.  Section~\ref{sec:measurements} lays out our graphical
results for Pauli measurements on stabilizer states and illustrates
the results with example measurements.  A detailed proof of the
measurement transformation is given in the Appendix.

\section{Background} \label{sec:background}

\subsection{Stabilizer-state graphs} \label{subsec:stabgraphs}

We now review the stabilizer-graph formalism presented in
Ref.~\cite{elliott:graphs}. Of central importance to the graphical
representation of stabilizer states is the fact that all stabilizer
states are equivalent under local Clifford operations to some graph
state~\cite{vandennest:complement,schlingemann:codes}. Furthermore,
the conversion of a graph state to any equivalent stabilizer state
can be achieved by applying to each qubit a single operation from the
following set: $I$, $Z$, $H$, $S$, $HZ$, and $SZ$ where the gates
$I$, $Z$, $H$, $S$ are the identity, sign-flip, Hadamard, and phase
gates, respectively.

As a consequence of these facts, one can draw a graph to represent
any stabilizer state by first drawing the graph corresponding to a
local-Clifford-equivalent graph state and then adding features
indicating which local gates must be applied to each qubit of the
graph state to transform it into the desired stabilizer state.

Simple graphs, those consisting of solid nodes connected by edges,
are used to represent graph states in the standard
way~\cite{hein:entanglement,vandennest:complement}.  In terms of a
preparation circuit, this amounts to associating with each node a
qubit initially prepared in the state $H\ket{0}$ and associating with
each edge a subsequent controlled-sign gate, $\CZ$, between the
qubits corresponding to the connected nodes.\footnote{$\CZ$ is often
called the controlled-phase or controlled-$Z$ gate.} Stabilizer-graph
notation augments this description by representing additional,
terminal gates as follows: the application of a $Z$ gate is
represented by a node with a negative sign, the application of an $S$
gate by a self loop, and the application an $H$ gate by a hollow
node.  If present, $H$ gates are assumed to act last; a hollow node
with a negative sign thus indicates the application of a $Z$ gate
followed by an $H$ gate.  In order to associate a stabilizer state
with any arrangement of solid and hollow nodes with arbitrary edges
and with or without self loops and signs, we choose to interpret a
hollow node with a loop as representing the application of an $S$
gate \emph{followed by\/} an $H$ gate. Although hollow nodes with
loops are not necessary to represent all stabilizer states, they are
a beneficial addition when considering the action of local Clifford
gates.\footnote{To represent all stabilizer states, it is also
unnecessary to include graphs in which any hollow nodes are connected
by edges (for discussion, see Ref.~\cite{elliott:graphs}), but we
considered such graphs in Ref.~\cite{elliott:graphs} and allow them
in our discussion in this paper. }

Given a stabilizer-state graph on $n$~nodes, we can easily write down
the associated $n$-qubit stabilizer state as follows.  Let
$\mathcal{H}$ denote the set of hollow nodes, $\mathcal{S}$ the set
of nodes with loops, and $\mathcal{Z}$ the set of nodes with negative
signs.  We label by $\Gamma$ the adjacency matrix of the underlying
graph, by which we mean the $n \times n$ square matrix whose entries
are determined from the graph by
\begin{equation}
\Gamma_{\!jk} = \left\{
\begin{array}{cl}
0\;, & \mbox{if $j = k$ or if $j\ne k$ are not connected,} \\
1\;, & \mbox{if $j \ne k$ are connected.}
\end{array} \right.
\end{equation}
The associated stabilizer state $\ket{\psi}$ is given by the formula
\begin{align} \label{eq:gatespsi}
\ket{\psi}&=
\prod_{m\in\mathcal{H}} H_m \prod_{l\in\mathcal{S}} S_l \prod_{k\in\mathcal{Z}} Z_k
\prod_{i,j}\left(\CZ_{ij}\right)^{\Gamma_{ij}} H^{\otimes n} \ket{0}^{\otimes n}\;,
\end{align}
where each gate acts upon the qubit(s) identified by its subscript(s).

As a final note, since qubits of a stabilizer state correspond to
nodes of a graph, we use the terms ``qubit'' and ``node''
interchangeably. Thus we speak of qubits in a graph and of applying
Clifford operations to nodes.

\subsection{Graph terminology} \label{subsec:terminology}

Much of what follows concerns the manipulation of stabilizer-state
graphs.  In preparation, this subsection introduces a variety of
terms describing graph transformations.  Some of these terms were
adopted from graph theory, while others have been invented for the
task at hand.

Among those terms common to graph theory are \emph{neighbors},
\emph{complement}, and \emph{local complement}.  The neighbors of a
node~$j$, which make a set denoted by $\set{N}(j)$, are those nodes
connected to $j$ by edges. In the results that follow, a loop does
not count as an edge, so a node is never its own neighbor.
Complementing the edge between two nodes removes the edge if one is
present and adds one otherwise. A local complement is performed by
complementing a selection of edges, with the pattern of edges
depending on whether local complementation is applied to a node or
along an edge.

\emph{Local complementation on a node\/} complements the edges
between all of the node's neighbors.  \emph{Local complementation
along an edge\/} is equivalent to a sequence of local
complementations on the nodes defining the edge. This sequence is as
follows: first perform local complementation on one of the nodes,
then local complement on the other node, and finally local complement
on the first node again. Local complementation along an edge is
symmetric in the two nodes defining the edge, so it does not matter
at which node local complementation is first performed.

To these terms we add \emph{flip\/} and \emph{advance}. Flip is used
to describe the simple reversal of some binary property, such as the
sign of a node or its fill, i.e., whether the node is solid or
hollow. Advance refers specifically to an action on loops; advancing
generates a loop on nodes where there was not previously one, and it
removes the loop and flips the sign on nodes where there was a loop.
Its action mirrors the application of the phase gate, since $S^2=Z$.

\subsection{Graphical description of Clifford operations}

We make use of the following transformation
rules~\cite{elliott:graphs}, which constitute a graphical description
of the action of $H$, $S$, and $Z$ gates on stabilizer states.
\begin{enumerate}
\item[T1.] Applying $H$ to a node flips its fill.
\item[T2.] Applying $S$ to a solid node advances its loop.
\item[T3.] Applying $S$ to a hollow node without a loop performs
local complementation on the node and advances the loops of its
neighbors.

If the node has a negative sign, flip the signs of its neighbors
as well.
\item[T4.] Applying $S$ to a hollow node with a loop flips its
fill, removes its loop, performs local complementation on it, and
advances the loops of its neighbors.

If the node does not have a negative sign, flip the signs of its
neighbors as well.
\item[T5.] Applying $Z$ to a solid node flips its sign.
\item[T6.] Applying $Z$ to a hollow node flips the signs of all of
its neighbors.  If the node has a loop, its own sign is flipped as
well.
\end{enumerate}

\subsection{Equivalent graphs}

Two graphs that look different can represent the same stabilizer
state. Because of this fact, we make use of the following equivalence
rules.
\begin{enumerate}
\item[E1.] Flip the fill of a node with a loop.  Perform local
complementation on the node, and advance the loops of its neighbors.

Flip the node's sign, and if the node now has a negative sign, flip
the signs of its neighbors as well.

\item[E2.] Flip the fills of two connected nodes without loops, and
local complement along the edge between them.

Flip the signs of nodes connected to both of the two original nodes.
If either of the two original nodes has a negative sign, flip it and
the signs of its current neighbors.
\end{enumerate}
Applying these equivalence rules to a stabilizer-state graph results
in a (generally different) graph that represents the same stabilizer
state. In fact, successive application of these rules generates all
graphs corresponding to a given state.

\section{Graphical description of Pauli measurements} \label{sec:measurements}

We now turn to our graphical formulation of Pauli measurements on
stabilizer states.

Let $M$ be an $n$-fold tensor product of the identity, $I$, and the
Pauli matrices, $X$, $Y$, and $Z$, that is,
\begin{equation}
M = \bigotimes_{j=1}^n M_j\;,
\end{equation}
where $M_j = I, X, Y,$ or $Z$.
If $M_j \ne I$, we call node~$j$ a
\emph{measured\/} node; otherwise, if $M_j = I$, we say the node is
not measured.

Given such a measurement operator, $M$, our task is twofold: first,
to find the probability that a measurement of $M$ on a quantum system
in the stabilizer state $\ket{\psi}$ gives an outcome ${(-1)}^a$,
and, second, to determine the post-measurement quantum state of the
system, i.e., a post-measurement stabilizer-state graph.   This
section describes a general graphical rule, applicable to the graph
that represents the stabilizer state, which accomplishes these tasks.

\subsection{Simplifying the measurement} \label{subsec:simplify}

By means of the graph transformation and equivalence rules reviewed in
Sec.~\ref{sec:background}, it is possible to greatly reduce the difficulty of formulating a
Pauli-measurement rule.  The following three
paragraphs describe a sequence of three simplifications that can be
made to any measurement, thereby restricting its form to one more
amenable to a measurement transformation rule.

The first simplification relies on the fact that a measurement where
$M_j = C Z C^{\dagger}$ is equivalent to a measurement where $M_j =
Z$ preceded by application of the local Clifford operation
$C_j^{\dagger}$ and followed by application of $C_j$ to the
post-measurement state.  If $M_j=X$, the local Clifford operation
needed is $C=H$; if $M_j=Y$, it is $C=SH$.  Thus, the first
simplification is to transform the original graph, using
rules~T1--T6, so that on the new graph the measurement becomes a
product of $Z$s on the measured nodes.  This means that it suffices
to determine the effect of $Z$-type measurements, that is,
measurements with the property that $M_j = I$ or $Z$ for all $j$. The
post-measurement state must be transformed by application of the
appropriate local unitaries to the measured nodes, i.e., $C_j$ to
measured node~$j$; in terms of graphs, this post-measurement
transformation is handled by rules~T1--T4.

The second simplification is to \emph{reduce\/} the graph, a
procedure introduced in Ref.~\cite{elliott:graphs}.  A reduced graph
is one in which hollow nodes are loopless and unconnected to one another.  Any stabilizer state can be represented
by a reduced graph.  Given a stabilizer state represented by a
stabilizer graph, one can find an equivalent reduced graph by
applying equivalence rules E$1$ and E$2$ to the given graph. One
applies rule E$1$ to any hollow node with a loop and uses E$2$ on any
pair of connected hollow nodes without loops.  Each application of
E$1$ or E$2$ makes solid the node(s) it is applied to, without
introducing new hollow nodes, so the procedure terminates in a
reduced graph. For the purposes of our measurement analysis, we are only
required to reduce the measured nodes, not the entire graph,
so the procedure terminates in a number of iterations that does not
exceed the number of measured nodes.  After this second
simplification, there are no loops on hollow measured nodes and no
edges between hollow measured nodes.

The final simplification is to disconnect hollow measured nodes from
unmeasured nodes by using equivalence rules E$1$ and E$2$.  Suppose
that a measured hollow node is connected to an unmeasured node.  In
the case that the unmeasured node does not have a loop, applying
equivalence rule E$2$ to the pair turns the measured node solid.  If
the unmeasured node has a loop, an application of E$1$ to the
unmeasured node gives the measured hollow node a loop.  Now one can
apply E$1$ to the measured hollow node to turn it solid.  One can
verify that in both cases, application of the equivalence rules
leaves the remaining measured hollow nodes loopless and unconnected
to one another.  Thus, this last simplification terminates in a
number of iterations no greater than the number of measured hollow
nodes.

The end product of these simplifications is a $Z$-type Pauli
measurement on a graph in which measured hollow nodes are loopless
and unconnected to one another and to unmeasured nodes.

\subsection{Graphical description of simplified measurements}
\label{subsec:gencase}

\begin{figure}
  \begin{tabular}{c@{}c@{}c@{}c@{}c@{}c@{}c@{}c@{}}
\includegraphics[width=2.1cm]{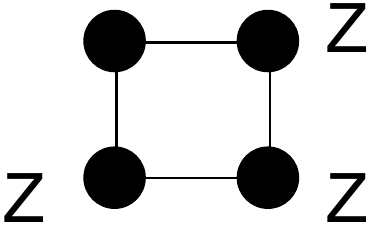} \hspace{.2em} & \raisebox{1.6em}{\Large $\stackrel{1}{\rightarrow}$} & \hspace{.3em} \raisebox{.15em}{\includegraphics[width=1.27cm]{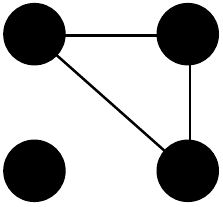}} \hspace{.3em} & \raisebox{1.6em}{\Large $\stackrel{2}{\rightarrow}$} & \hspace{.3em} \raisebox{.15em}{\includegraphics[width=1.25cm]{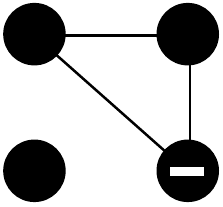}} \hspace{.3em} & \raisebox{1.6em}{\Large$\stackrel{3}{\rightarrow}$} & \hspace{.3em} \raisebox{.15em}{\includegraphics[width=1.25cm]{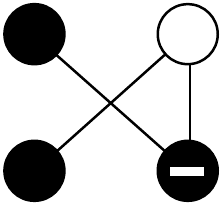}}
  \end{tabular}
\caption{The steps that perform a Pauli product measurement $M = I
\otimes Z \otimes Z \otimes Z$ on a stabilizer-state graph for the
four-qubit cluster state.  The juxtaposition of Pauli operator and
node indicates the presence of that operator in the intended
measurement.  We label the nodes in the graphs clockwise starting
from~$1$ in the upper left corner.  Thus, $\setM = \{ 2,3,4 \}$,
$\smallsetMS = \smallsetMSE = \{ 2,3,4 \}$, and $\smallsetMH =
\varnothing$.  Since $\smallsetMSE \ne \varnothing$, the outcome is
random; for the sake of illustration, we take it to be $+1$. Node~$2$
is taken to be the chosen node, so the edges between its neighbors,
nodes $\{ 1,3 \}$, and the unchosen nodes in $\smallsetMSE$, nodes
$\{ 3,4 \}$, are complemented in step~$1$. The chosen node does not
have a sign, and $a+b$ is even if we assume a $+1$ measurement
outcome, so step~$2$ has the effect of giving a sign to the node in
$\setMSE$ that is also a neighbor of node $2$, meaning node $3$.
Step~3 removes all edges involving node~$2$ and then connects it to
nodes~$3$ and~$4$ while making node~$2$ hollow.  Finally, step~$4$
has no effect since the chosen node has no loop.  If either
node~$3$ or $4$ had been picked as the chosen node, the resulting
graph could be transformed into this one by using equivalence rule E$2$.
\label{fig:ZZZZexample}}
\end{figure}

With the preceding simplifications carried out, we can now spell out
the graphical description of the measurement.  The proof of this
description is given in terms of circuit identities in the Appendix.

In addition to the sets $\setH$, $\setS$, and $\setZ$ introduced in
Sec.~\ref{subsec:stabgraphs}, we require the use of several other
sets of nodes in the stabilizer-state graph: $\setM=\{j\mid M_j\ne
I\}$ is the set of measured nodes, $\setMS=\setM\backslash \setH$ is
the set of measured solid nodes, $\setMH=\setM\cap\setH$ is the set
of measured hollow nodes, and
$\setMSE=\{j\in\setMS\mid|\setMH\cap\set{N}(j)|=\mbox{$0\pmod2$}\}$
is the set of measured solid nodes that have an even number of
connections to measured hollow nodes.  Here
$\set{A}\backslash\set{B}$ denotes the set of elements in $\set{A}$
that are not in $\set{B}$, $\set{A}\cap\set{B}$ denotes the
intersection of $\set{A}$ and $\set{B}$, and $|\set{A}|$ denotes the
number of elements in $\set{A}$.

When a Pauli measurement is made on a stabilizer state, the outcome
is either random, with the two possible outcomes being equiprobable, or certain.  Which case applies depends on $\setMSE$.  The
outcome in the deterministic case is specified by
\begin{equation}\label{eq:b}
b=|\setMH\cap\setZ|\;,
\end{equation}
the number of measured hollow nodes with a sign.

The result of a $Z$-type Pauli measurement is as follows.
\begin{enumerate}
\item If $\setMSE = \varnothing$, the measurement outcome is
${(-1)}^b$ with certainty, and the state is unchanged by the
measurement.
\item If $\setMSE \ne \varnothing$, the measurement outcome,
$(-1)^a$, is random, and a graph for the post-measurement state
can be obtained according to steps $1$--$4$ below.
\end{enumerate}

To find the post-measurement state when $\setMSE \ne \varnothing$, it
is necessary first to pick a node, which we call the \emph{chosen
node}, from $\setMSE$.  The post-measurement state is then obtained
by the following four steps.
\begin{enumerate}
\item For each neighbor of the chosen node, complement all
of its edges to unchosen nodes in $\setMSE$.
\item If the chosen node has no sign, flip the signs
of all its neighbors that are also in $\setMSE$; otherwise, if
the chosen node has a sign, remove that sign, and flip the signs
of all other nodes in $\setMSE$ that do not neighbor the chosen
node. If $a+b$ is odd, flip the signs of the chosen node and all
its neighbors.
\item Remove all edges involving the chosen node, and then connect
the chosen node to all the other nodes in $\setMSE$.  Make the
chosen node hollow.
\item If the chosen node has a loop, remove that loop, perform
local complementation on the chosen node, advance the loops of
its neighbors, and if $a+b$ is odd, flip the signs of the
unchosen nodes in $\setMSE$.
\end{enumerate}
These steps constitute a complete graphical description for the
effect of the measurement $M$ on the state.  Notice that, in
step~$1$, an edge between two nodes that are in $\setMSE$ and are
initially neighbors of the chosen node gets complemented twice, so it
remains unchanged.  Figure~\ref{fig:ZZZZexample} illustrates the use
of these rules for the case of a three-qubit measurement on a
four-qubit cluster state.

In the case that $\setMSE$ has a single element, it becomes the
chosen node, and the steps in the measurement transformation rule
simplify to the following.
\begin{enumerate}
\item[$1'$.]Do nothing.  There are no unchosen nodes in $\setMSE$.
\item[$2'$.]If the chosen node has a sign, remove that sign.
If $a+b$ is odd, flip the signs of the chosen node and all of its
neighbors.
\item[$3'$.]Disconnect the chosen node from the graph, and make it hollow.
\item[$4'$.]If the chosen node has a loop, remove that loop.
\end{enumerate}
These four steps can be summarized as follows:  Remove all loops and
signs from the chosen node.  Flip the signs of the chosen node and
its neighbors if $a+b$ is odd.  Then disconnect the chosen node from
the graph, and make it hollow.

\subsection{Single-qubit measurements}
\label{subsec:singlequbit}

\begin{figure*}
  \begin{tabular}{lc@{}c@{}c@{}c@{}c@{}c@{}c@{}c@{}c}
    \raisebox{4.7em}{(a)\hspace{0em}} & \includegraphics[width=2.8cm]{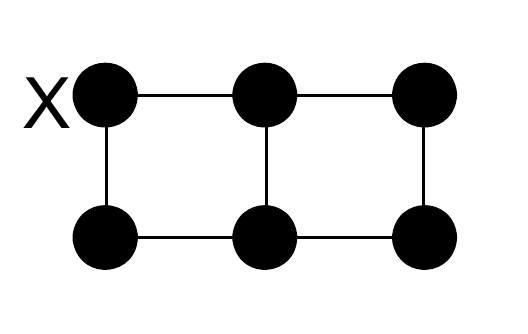} & \raisebox{2.4em}{\Large $\stackrel{H}{\rightarrow}$} & \includegraphics[width=2.8cm]{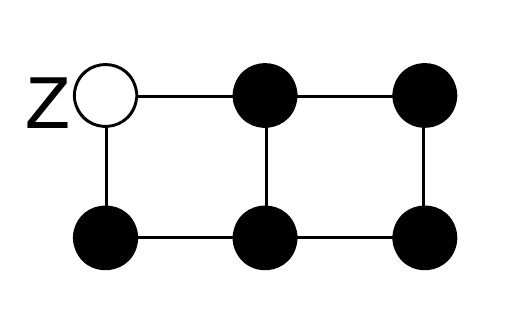} & \raisebox{2.4em}{\Large $\cong$} & \includegraphics[width=2.8cm]{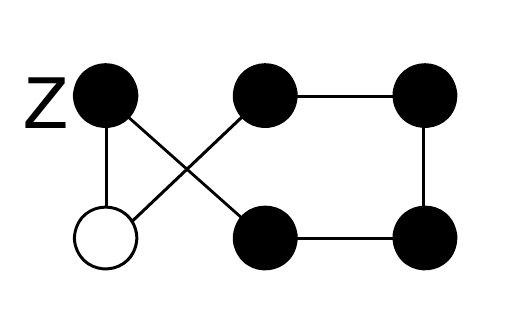} & \raisebox{2.4em}{\Large$\stackrel{P_0}{\longrightarrow}$} & \includegraphics[width=2.8cm]{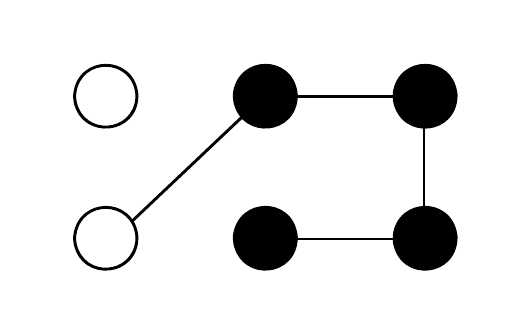} & \raisebox{2.4em}{\Large $\stackrel{H}{\rightarrow}$} & \includegraphics[width=2.8cm]{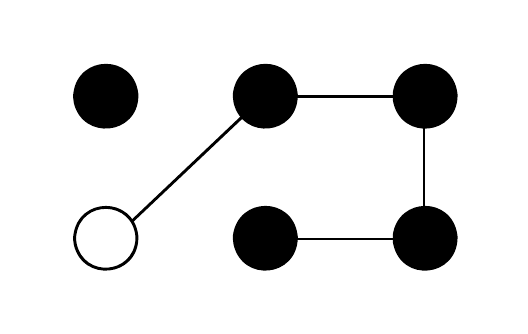} \\
    \raisebox{4.7em}{(b)\hspace{0em}} & \includegraphics[width=2.8cm]{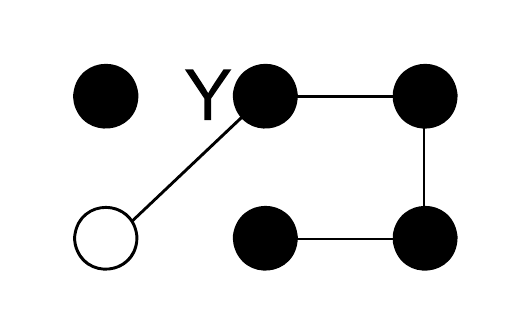} & \raisebox{2.4em}{\Large $\stackrel{HS^\dagger}{\longrightarrow}$} & \includegraphics[width=2.8cm]{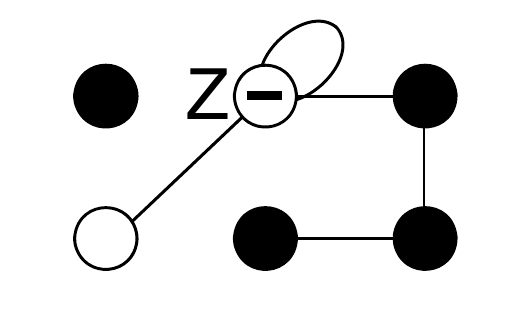} & \raisebox{2.4em}{\Large $\cong$} & \includegraphics[width=2.8cm]{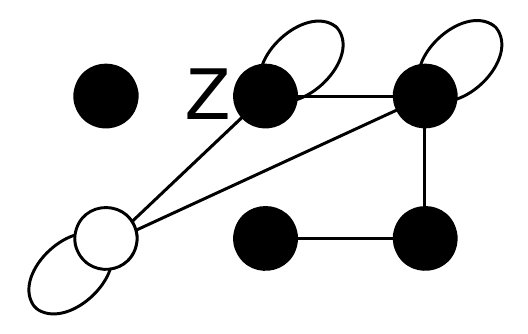} & \raisebox{2.4em}{\Large$\stackrel{P_0}{\longrightarrow}$} & \includegraphics[width=2.8cm]{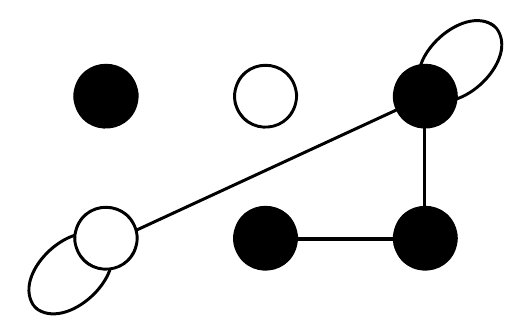} & \raisebox{2.4em}{\Large $\stackrel{SH}{\longrightarrow}$} & \includegraphics[width=2.8cm]{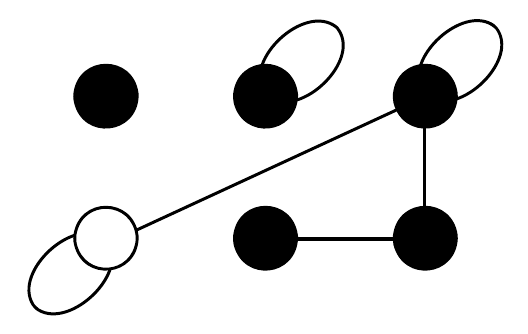} \\
    \raisebox{4.7em}{(c)\hspace{0em}} & \includegraphics[width=2.8cm]{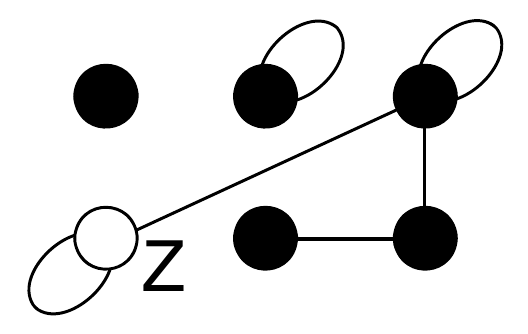} & \raisebox{2.4em}{\Large $\cong$} & \includegraphics[width=2.8cm]{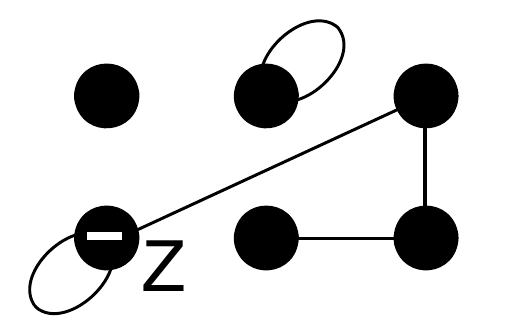} & \raisebox{2.4em}{\Large$\stackrel{P_0}{\longrightarrow}$} & \includegraphics[width=2.8cm]{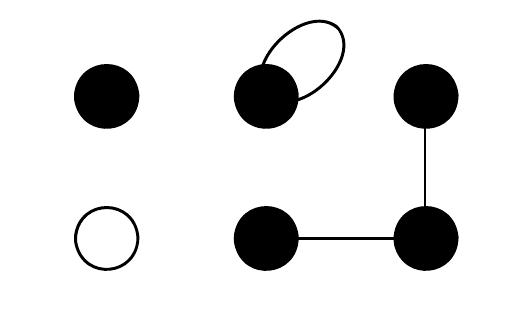}
  \end{tabular}
\caption{Examples of the graph manipulations associated with
successive single-qubit measurements of (a)~$X$, (b)~$Y$, and (c)~$Z$
on a $2\times 3$ cluster state.  The juxtaposition of Pauli operator
and node is used to indicate the intended measurement of that
operator on the node. State transformations are indicated by arrows
labeled by the transformation being applied to the measured qubit.
All $Z$ measurements are assumed to yield outcome $+1$, thereby
applying the projector $P_0$ to the measured qubit.  Let the nodes be
labeled clockwise starting from $1$ in the upper left corner. The
measurement of $X_1$ in (a) is accomplished by transforming both
state and measurement by $H_1$, applying equivalence rule E$2$ to
nodes $1$ and $6$, applying the measurement transformation to node
$1$, and, finally, applying $H_1$ to the resultant state. Similarly,
the measurement of $Y_2$ in (b) is accomplished by transforming both
state and measurement by $H_2S^\dagger_2$, applying equivalence rule
E$1$ to node $2$, applying the measurement transformation to node
$2$, and applying $S_2H_2$ to the resultant state. The measurement of
$Z_6$ in (c) requires only the application of equivalence rule E$1$
to node $6$, followed by application of the measurement
transformation to node $6$.  In each case the equivalence rule is
necessary to fill the node of interest so that our measurement rule
for $Z$ can be applied.  In parts (a) and (b) it is first necessary
to transform the measurement to a $Z$ measurement.  By transforming
the state as well, we leave the measurement outcome invariant and
ensure that the inverse transformation applied after the $Z$
measurement yields the appropriate state. (This caption is a tribute
to the late John A. Wheeler and his devotion to writing long,
self-explanatory captions.) \label{fig:example}}
\end{figure*}

Single-qubit measurements are a straightforward but important special
case~\cite{aaronson:simulate}, as illustrated, for example, by the
use of such measurements in measurement-based quantum
computation~\cite{raussendorf:oneway}.  In this section, we
specialize the measurement transformation rule of the previous
subsection to Pauli measurements on a single measured node, thereby
producing a very simple graphical description of such single-qubit
measurements.

In the context of single-qubit measurements, the simplification
procedure in Sec.~\ref{subsec:simplify} ensures that $\setM$ consists
either of a solid node or of a loopless hollow node that is
disconnected from the rest of the graph.  In the latter case, the
measured qubit is in the state $|0\rangle$ if it has no sign or
$|1\rangle$ if it does; the outcome is certain
($\setMSE=\varnothing$) and equal to the sign of the hollow node. In
the former case, $\setMSE$ consists of the measured solid node,
$b=0$, and the outcome $(-1)^a$ is random.  In this case, steps~1--4
of Sec.~\ref{subsec:gencase}, which generate a post-measurement
graph, reduce to steps~$1'$--$4'$ with $b=0$.  Thus, the four steps
now reduce to the following: Remove all loops and signs from the
measured node.  Flip the signs of the measured node and its neighbors
if the outcome is $-1$ ($a$ is odd). Then disconnect the chosen node
from the graph, and make it hollow.

We illustrate the transformations associated with single-qubit
measurements in Fig.~\ref{fig:example}.  To highlight the relevance
of our results to measurement-based quantum computation, successive
$X$, $Y$, and $Z$ measurements on a $2\times3$ cluster state are
considered.  We purposefully follow an inefficient measurement order
for pedagogical reasons.  In the examples, each measurement outcome
is random, and, for purposes of constructing a post-measurement
graph, we assume the outcome of each measurement is $+1$.

\section{Conclusion}

In this paper we have presented a graphical rule for transforming
stabilizer states under the measurement of products of Pauli
operators both in the general case and in the special case of
single-qubit measurements.  Together with the transformation rules
for Clifford gates given in Ref.~\cite{elliott:graphs}, the
transformation rules for Pauli measurements allow any operation
taking one stabilizer state to another to be represented pictorially.
Thus, the present paper completes a novel graphical representation of
these ubiquitous states.

\acknowledgments
Thanks to Adam Meier, Manny Knill, and David Smith for their careful readings of this document.
The work presented here was supported in part by National Science
Foundation Grant No.~PHY-0653596.  Contributions by NIST, an agency of the US government, are not subject to copyright laws.

\appendix

\section*{Appendix: Proof of the general case} \label{app:proof}

\begin{figure*}
\center
\includegraphics[width=18cm]{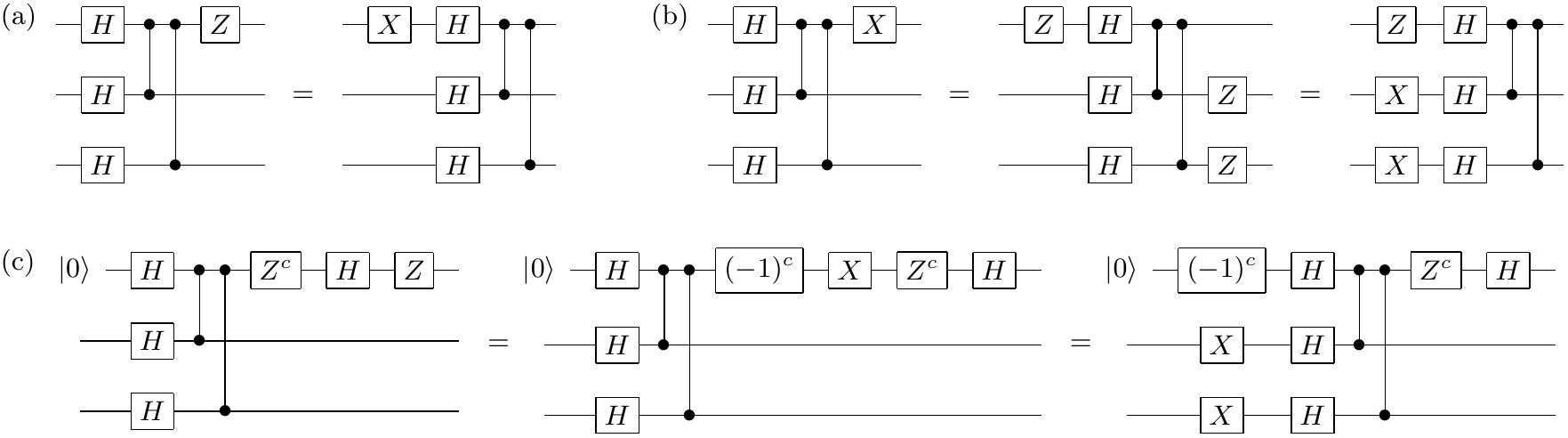}
\caption{Circuit identities used to determine the effect of a $Z$
operator on a stabilizer state.  Identity~(a) follows from the fact
that $Z$ and controlled-sign gates commute and from the identity $HZH
= X$.  A similar identity holds if the node has a loop or a sign
since $Z$ commutes with itself and with $S$.  The first equality in
identity~(b) is easily verified in the standard basis, and the second
equality is an application of~(a).  Identity~(c) follows from the
equalities shown; the second equality uses the identity in~(b).
\label{fig:Zoperators}}
\end{figure*}

In this Appendix we derive the measurement rule given in
Sec.~\ref{subsec:gencase} for transforming a stabilizer graph under a
measurement $M$, where $M$ is any tensor product of $I$ and $Z$ Pauli
operators.  The proof proceeds in three stages.  In the first, we
determine the effect of $M$, considered as a Clifford unitary
operation, on a stabilizer state $\ket\psi$.  This enables us, in the
second stage, to find the action of the measurement projector $P_a
=[I+{(-1)}^a M]/2$ on $\ket\psi$ and thus to determine whether the
measurement is certain or random.  The post-measurement state is then
found via a simple circuit identity in the last stage.  Notice that,
in determining the effect of $M$ on $\ket\psi$, we must retain the
overall phase, since, in the second stage, the overall phase becomes
a relative phase in the superposition $[\ket\psi+(-1)^a
M\ket\psi]/2$.

\subsection{Action of $Z$ on a stabilizer state}

To begin, consider the effect of a unitary $M$ on the $n$-qubit
stabilizer state $\ket{\psi}$, where $M$ is known to be a tensor
product consisting of only the operators $I$ and $Z$.  The action of
$I$ is trivial, so we can focus on determining the action of $Z$.  As
illustrated in Fig.~\ref{fig:Zoperators}(a), applying $Z$ to a solid
node is equivalent to applying an $X$ operator to the same node prior
to all other Clifford gates in the circuit.  Similarly, the action of
$Z$ on a hollow node can be reexpressed using the circuit identity in
Fig.~\ref{fig:Zoperators}(c).  This identity shows that an identical
state is obtained by adding a leading $X$ operator to each neighbor
of the hollow node and, if the hollow node has a sign, introducing an
overall phase of $-1$. For properly simplified measurements, hollow
measured nodes are neighbored only by solid measured nodes, so only
members of $\setMS$, the measured solid nodes, collect leading $X$
operators.  The number of $X$ operators collected by each member of
$\setMS$ is $1$ plus the number of neighbors it has in the set
$\setMH$.  Since $X^2=I$, the net result is that $X$ operators are
added only to members of $\setMSE$, the set of solid measured nodes
with an even number of hollow measured neighbors.

Summarizing, the stabilizer state $M\ket{\psi}$ can be obtained from
the state $\ket{0}^{\otimes n}$ by first applying an $X$ to each
member of $\setMSE$ and then applying the Clifford gates needed to
obtain $\ket{\psi}$ from the initial state $\ket{0}^{\otimes n}$.  In
addition, an overall phase of $(-1)^b$ must be applied, where
$b=|\setMH\cap\setZ|$ is the number of measured hollow nodes with a
sign.  That is,
\begin{equation}
M\ket\psi=
MU\ket{0}^{\otimes n}=(-1)^b
U\Biggl(\prod_{j\in\smallsetMSE}X_j\Biggr)\ket{0}^{\otimes n}\;,
\end{equation}
where $U$ is shorthand for the sequence of Clifford gates for
preparing $\ket{\psi}=U\ket{0}^{\otimes n}$, as indicated by the
stabilizer graph (see Eq.~(\ref{eq:gatespsi})).

\begin{figure*}
\center
\includegraphics[width=18cm]{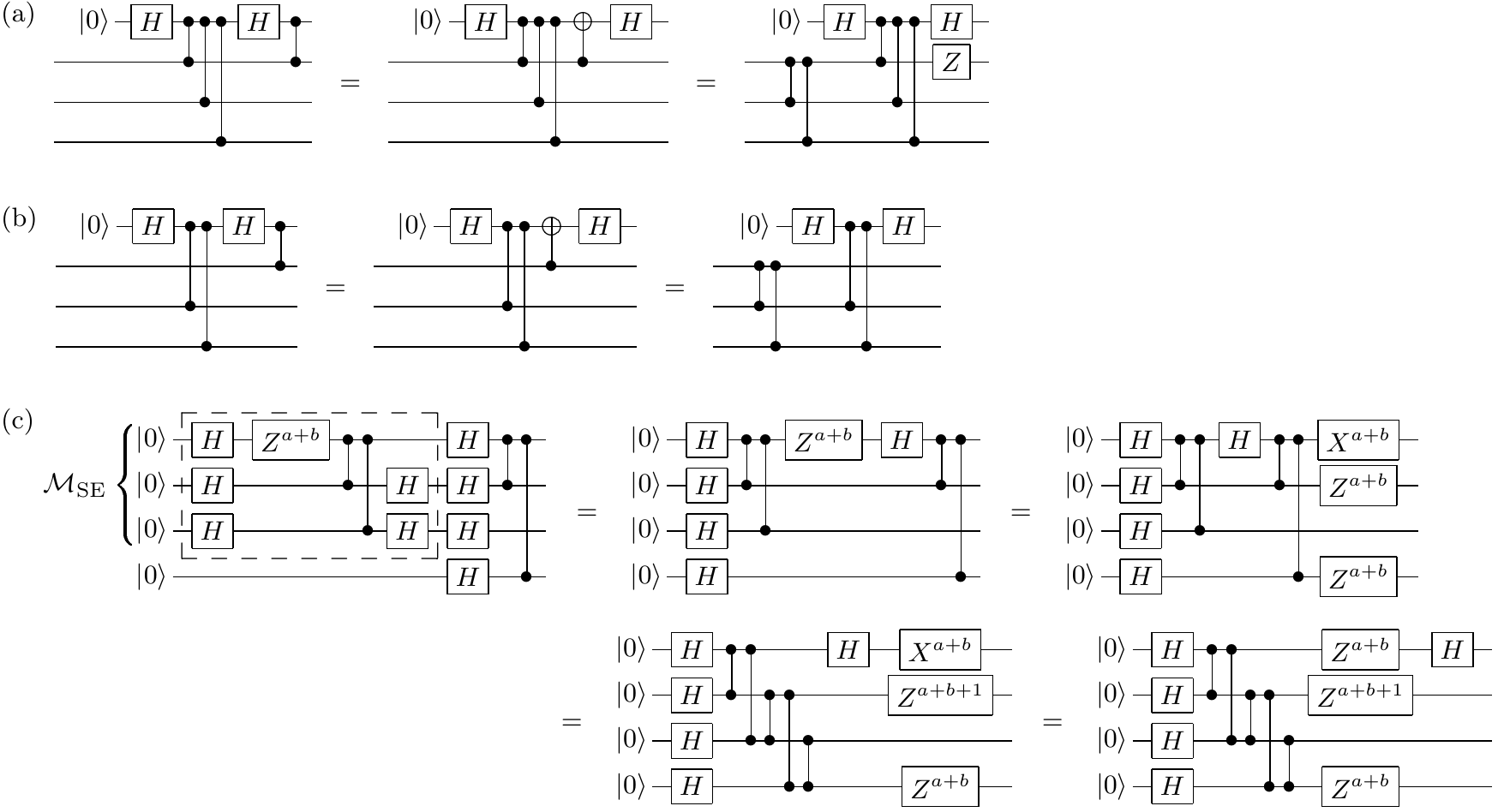}
\caption{Circuit identities used to determine the post-measurement
state. Identities~(a) and~(b) utilize basic circuit identities for
pushing a controlled-not gate past a controlled-sign gate.  For the
identity in~(c), the first three qubits represent the elements of
$\smallsetMSE$, with the chosen qubit $p$ placed on top.  The dashed
box delineates the Clifford operations that put the qubits in
$\smallsetMSE$ into an appropriate cat state, i.e., those operations
in Eq.~(\ref{eq:poststate}) that are applied before $U$, whereas the
gates after the dashed box are the relevant portion of the gates in
$U$, i.e., the gates that create the original stabilizer state. The
first equality in~(c) is trivial.  The second uses the identity in
Fig.~\ref{fig:Zoperators}(b) to push $Z^{a+b}$ to the far right of
the circuit, turning it into $X^{a+b}$ and depositing $Z^{a+b}$ on
each neighbor of $p$.  The next equality eliminates the
controlled-sign gates that initially connected $p$ to other nodes by
using the identity in~(a) for connections to nodes in $\smallsetMSE$
and using~(b) for connections to nodes outside of $\smallsetMSE$. The
final simple equality returns the circuit to graph form.
\label{fig:catcircuit}}
\end{figure*}

\subsection{Certain and random outcomes}

The second stage of our proof applies this result to find the action
of $P_a$ on $\ket\psi$ and the probability of obtaining measurement
outcome ${(-1)}^a$, which is given by $\bra{\psi} P_a \ket{\psi}$.
We have immediately that
\begin{align}
P_a \ket{\psi} &= \frac{1}{2}[I+{(-1)}^a M]\ket\psi\\
&=U\frac{1}{2}\Biggl(I+(-1)^{a+b}
\prod_{j\in\smallsetMSE}X_j\Biggr)\ket{0}^{\otimes n}\;,
\end{align}
which gives
\begin{align}
\bra{\psi} P_a &\ket{\psi}
= \frac{1}{2}\!\left(1 + (-1)^{a+b}\bra{0}^{\otimes n}
\Biggl(\prod_{j\in\smallsetMSE}X_j\Biggr)\ket{0}^{\otimes n}\right)\;.
\end{align}
Since $\bra{0} X \ket{0} = 0$, this means that measurements are
classified into two types: if $\setMSE = \varnothing$, the outcome
probabilities are $1$, for $a=\mbox{$b\pmod2$}$, and $0$, for $a \ne
\mbox{$b\pmod2$}$, but if $\setMSE \ne \varnothing$, $a$ has a $50\%$
chance of being either $0$ or $1$.

\subsection{Post-measurement state}

When the measurement outcome is ${(-1)}^a$, the post-measurement
state is $\ket{\psi'}=P_a\ket{\psi}/\sqrt{\bra{\psi}P_a\ket{\psi}}$.
When $\setMSE = \varnothing$, the outcome $a=b$ occurs with
certainty, and the post-measurement state is the same as the original
stabilizer state $\ket\psi$.

When $\setMSE \ne \varnothing$, the situation is more complicated. In
this case $\bra{\psi}P_a\ket{\psi}=1/2$, so
\begin{equation}
\ket{\psi'} =
U\frac{1}{\sqrt{2}}\Biggl(I+{(-1)}^{a+b}
\prod_{j\in\smallsetMSE}X_j\Biggr)\ket{0}^{\otimes n}\;.
\end{equation}
Thus the post-measurement state is obtained by use of the Clifford
circuit that created the original stabilizer state, but applied to an
initial state that, for the qubits in $\setMSE$, is changed to a cat
state, i.e., an equal superposition of all 0s and all 1s, with the
sign between the two terms in the superposition given by
$(-1)^{a+b}$.  To construct a graph for the post-measurement state,
we introduce a standard quantum circuit for making the cat state from
$\ket{0}^{\otimes n}$ and use circuit identities to put the overall
quantum circuit into graph form. Thus we write
\begin{align} \label{eq:poststate}
\ket{\psi'}=
U\Biggl(\prod_{l\in\smallsetMSE\backslash p}&H_l\Biggr)
\Biggl(\prod_{k \in\smallsetMSE\backslash p}\CZ_{pk}\Biggr)\nonumber\\
&\times Z_p^{a+b}\prod_{j\in\smallsetMSE}H_j\ket{0}^{\otimes n}\;,
\end{align}
where $p$ denotes the chosen node.  Figure~\ref{fig:catcircuit}
translates these Clifford operations into circuit notation and
develops the identities needed to determine the post-measurement
state $\ket{\psi'}$.

Most of the components of steps $1$--$4$ in Sec.~\ref{subsec:gencase}
follow directly from Fig.~\ref{fig:catcircuit}(c), which begins with
a circuit for a representative post-measurement state that is not in
graph form.  The boxed portion of the circuit puts the qubits in
$\setMSE$ into a cat state; that which follows is the relevant
portion of the graph-form circuit for the pre-measurement state.
Application of a sequence of identities yields the circuit on the far
right, which is in the proper form to translate to a stabilizer-state
graph. From the first equality in Fig.~\ref{fig:catcircuit}(c), it
can be seen that the new connections specified by rule~$3$ between
$p$ and the unchosen nodes in $\setMSE$ arise directly from the
prepended cat state. The sign changes of $p$'s neighbors given at the
end of rule~$2$ follow from second equality.  Pushing the Hadamard
right of the remaining $\CZ$ gates in the third equality and pushing
the resultant $\CX$ gates the other way removes $p$'s previous
connections while complementing edges between nodes that were
neighbors of $p$ and the unchosen nodes in $\setMSE$; these
operations appear in rules~$3$ and $1$, respectively.  In the same
equality, nodes that are both neighbors of $p$ and elements of the
set $\setMSE$ pick up the additional sign specified in the initial
part of rule~$2$. In the final equality, the chosen node becomes
hollow, as specified in rule~$3$ and adopts a final sign of
$(-1)^{a+b}$ as specified in rule $2$.

The final components of the measurement transformation rules deal
with the cases in which the chosen node initially has a sign and/or a
loop. The circuit identity in Fig.~\ref{fig:catcircuit}(c) does not
explicitly cover these situations, but they are easily derived by
applying the gates in question to either end of the identity and then
applying the appropriate transformation rules to the right side.
Rule~$4$ follows from transformation rule~T$3$, while the alternative
sign change in the initial part of rule $2$ arises from
transformation rule~T$6$.

\end{document}